\documentclass{nature}
\usepackage{lettrine}
\usepackage{graphicx}
\usepackage{dcolumn}
\usepackage{bm}
\usepackage{amssymb}
\usepackage{amssymb}
\usepackage{epsfig}
\usepackage[10pt]{moresize}
\usepackage{comment}
\usepackage{graphicx}
\usepackage{epsfig}
\usepackage{subfigure}
\usepackage{epstopdf}
\usepackage{cite}
\usepackage{epstopdf}
\usepackage{amsfonts}
\usepackage{amsmath}
\usepackage{amstext}
\usepackage{esint}
\usepackage{graphicx}
\usepackage{graphics}
\usepackage{float}
\usepackage{epsfig}
\usepackage{subfigure}
\usepackage{multirow}

\begin{document}

\title{Experimental simulation of quantum temporal steering beyond
rotating-wave approximation}
\author{Shao-Jie Xiong$^{1,2}$, Yu Zhang$^{1}$, Zhe Sun$^{1\star}$, Li Yu$%
^{1}$, Jinshuang Jin$^{1}$, Xiao-Qiang Xu$^{1}$, Jin-Ming Liu$^{2}$, Kefei
Chen$^{3}$, Chui-Ping Yang$^{1\dag}$}
\maketitle

\begin{abstract}
Characterizing the dynamics of open systems usually starts with a
perturbative theory and involves various approximations, such as the Born,
Markov and rotating-wave approximation (RWA). However, the approximation
approaches could introduce more or less incompleteness in describing the
bath behaviors. Here, we consider a quantum channel, which is modeled by a
qubit (a two-level system) interacting with a bosonic bath. Unlike the
traditional works, we experimentally simulate the system-bath interaction
without applying the Born, Markov, and rotating-wave approximations. To our
knowledge, this is the first experimental simulation of the quantum channels
without any approximations mentioned above, by using linear optical devices.
The results are quite useful and interesting, which not only reveal the
effect of the counter-rotating terms but also present a more accurate
picture of the quantum channel dynamics. Besides, we experimentally
investigate the dynamics of the quantum temporal steering (TS), i.e., a
temporal analogue of Einstein-Podolsky-Rosen steering. The experimental and
theoretical results are in good agreement and show that the counter-rotating
terms significantly influence the TS dynamics. There are obviously different
dynamics of TS in non-RWA and RWA channels. However, we emphasize that the
results without RWA are closer to realistic situations and thus more
reliable. Due to the close relationship between TS and the security of the
quantum cryptographic protocols, our findings are expected to have useful
applications in secure quantum communications and future interesting TS
studies.
\end{abstract}

\begin{affiliations}
\item Department of Physics, Hangzhou Normal University,
Hangzhou, Zhejiang 310036, China
\item State Key Laboratory of Precision Spectroscopy,
Department of Physics, East China Normal University, Shanghai
200062, China
\item Department of Mathematics, Hangzhou Normal University, Hangzhou 310036, China
\\$^\star$e-mail:sunzhe@hznu.edu.cn
\\$^\dag$e-mail:yangcp@hznu.edu.cn
\end{affiliations}
%\date{\today}

%\pacs{03.67.Bg, 42.50.Dv, 85.25.Cp, 76.30.Mi}
%\pacs{03.67.Bg, 42.50.Dv, 85.25.Cp, 76.30.Mi}

Einstein-Podolsky-Rosen (EPR) steering is one of the most essential features
in quantum mechanics. For a bipartite system in an entangled state, EPR
steering problems refer to the quantum nonlocal correlations, which allow
one of the subsystems to remotely prepare or steer the other one via local
measurements. EPR steering is ususlly treated as an intermediate scenario
lying in between the entanglement and the Bell nonlocality. As a result,
some entangled states cannot be employed to achieve steering tasks, and some
steerable states do not violate Bell-like inequalities~\cite{1,2}. Recently,
quantum steering problems have attracted considerable interest\thinspace
\cite{3,4,5,5-1,6,7,8,8-1,9,10,11,12,13,14,15,16,17,18,19,20}. Apart from
the fundamental interests of steering in quantum mechanics~\cite%
{1,2,3,4,5,5-1,6,7,8,8-1,9,10}, there are lots of application motivations
for EPR steering which is thought to be the driving power of quantum secure
communication and teleportation\thinspace \cite{11,12,13}. For example,
steering allows quantum key distribution (QKD) if one party trusts his own
devices but not those of the other party\thinspace \cite{12}. An outstanding
advantage is that this steering-dominated scenario is easier to implement,
when compared with the completely-device-independent protocols\thinspace
\cite{13}.

Several inequalities in terms of sufficient conditions of steerability were
employed to detect steerable states\thinspace \cite{1,2,3,4}, and such
inequalities have been tested by several experiments~\cite{5,5-1,6,7,8}.
Beyond inequalities, a number of possible measures were developed to
quantify the quantum steerability \cite{14,15}. Very recently, Skrzypczyk
et\thinspace al. proposed a powerful measure named the steerable weight~\cite%
{16,17}. Moreover, steerability was found to be equivalent to joint
measurability~\cite{18,19}. A close relationship between steerability and
quantum-subchannel discrimination was discovered in~\cite{20}.

Among the steering studies, a novel and important direction is to consider
quantum steering in time, i.e., the so-called temporal steering (TS)~\cite%
{21}. Different from discussing the spatially-separated systems, TS problems
focus on a single system at different times. In this frame, a system is sent
to a distant receiver (say Bob) through a quantum channel, then a detector
or manipulator (say Alice) performs some operations (including measurements
on the system) before Bob receives the system and performs his measurement.
The nonzero TS accounts for how strongly Alice's choice of measurements at
an initial time can influence the final state captured by Bob. In addition,
TS also reveals a unique link between a quantum system's past and future
features. No third party can steer as strongly as Bob or gather full
information about the final state. Based on this quantum origin, the TS
inequality\thinspace \cite{21,21-0,21-1} becomes a very useful tool in
verifying the suitability of a quantum channel for a certain QKD process.
Compared with the spatial steering inequalities, TS inequality presents a
superior applicability. Thus, the entanglement based scenarios are no longer
necessary, and one can directly carry out the famous BB84 protocol and other
related protocols\thinspace \cite{12,21,21-1,21-2}.

In order to quantify TS, a concept of \textit{temporal steerable weight} was
introduced in the literature\thinspace \cite{22}, where the authors found
that TS characterized by the weight can be used to define a sufficient and
practical measure of strong non-Markovianity. This implies the evident
dependence of TS dynamics on the properties of quantum channels. Moreover,
it was found that TS is intrinsically associated with realism and joint
measurability~\cite{23,24}.

In this paper, we investigate the influence of a quantum channel on the TS
behavior. This is a particular TS problem rather than EPR steering, because
the formation of the quantum correlation between the system's initial and
final state lies on the quantum channel. For example, during a unitary
evolution, Alice is able to perfectly steer the final state received by Bob.
A nontrivial case of TS refers to a non-unitary dynamics when Alice's
influence is erased partially or completely. Because TS is sensitive to the
characteristics of the channel, a natural question arises: what kind of
channel should we take into account?

Usually, the description of the dynamics of open systems starts with a
perturbative theory and involves various approximations, such as the Born,
Markov and rotating-wave approximation~(RWA). However, it is widely believed
that the approximation approaches introduce more or less the incompleteness
of the description of the bath. An efficient numerical method that avoids
using the above approximations was developed by Tanimura et~al.~\cite{25,26}%
, who established a set of hierarchical equations that includes all orders
of system-bath interactions. The hierarchy equation method has been
successfully employed to describe the quantum dynamics of various physical
and chemical systems~\cite{27,28} as well as some quantum devices\thinspace
\cite{29,30}.

The aims of this paper are as follows. Firstly, we propose a linear-optical
setup to experimentally simulate a quantum channel modeled by a quantum
two-level system (i.e., qubit) interacting with a bosonic bath, without
applying Born, Markov and rotating-wave approximations. The experimental
parameters accounting for the dynamics of the qubit are set by means of the
hierarchy equation method. To the best of our knowledge, this is the first
experimental simulation of this kind of quantum channel in a linear optical
setup. Our experimental simulation offers a fruitful testbed for the studies
of open quantum systems. Secondly, we experimentally investigate the TS
problems in this channel. Our study indeed provides a more realistic
representation of the environment influence than the usual quantum channels
such as the amplitude damping channel, where the environment was only
treated simply. Moreover, we found that there is a lack of experimental
study of TS problems up to now. Some existing study~\cite{31} was only based
on a phenomenologically-designed channel and thus it is impossible to
highlight the important role of the system-bath interaction. In contrast,
the channel under our study, governed by a system-bath Hamiltonian without
RWA, allows us not only to reflect the special roles of the counter-rotating
terms, but also to consider the system-bath couplings in arbitrarily strong
regimes.

\section*{Results}

\textbf{System-bath model. }We consider a qubit system interacting with a
bosonic bath, described by a full Hamiltonian:
\begin{equation}
H=H_{\text{S}}+H_{\text{B}}+H_{\text{Int}},  \label{eq:Ham}
\end{equation}%
where $H_{\text{S}}=\frac{\omega _{0}}{2}\sigma _{z}$ is the free
Hamiltonian of the qubit (assuming $\hbar =1$), with $\sigma _{z}$ being the
Pauli operator of the qubit and $\omega _{0}$ standing for the transition
frequency between the two levels of the qubit; $H_{\text{B}}=\sum_{k}\omega
_{k}b_{k}^{\dagger }b_{k}$ is the free Hamitonian of the bosonic bath with $%
b_{k}^{\dagger }$ and $b_{k}$ being the bosonic creation and annihilation
operators of the $k$th mode of frequency $\omega _{k}$; and
\begin{equation}
H_{\text{Int}}=\sum_{k}\sigma _{x}\left( g_{k}b_{k}+g_{k}^{\ast
}b_{k}^{\dagger }\right)  \label{non-RWA-Hint}
\end{equation}%
is the interaction Hamiltonian between the qubit and the bath with $g_{k}$
being the coupling strength between the qubit and the $k$th mode of the
bath. One essential aspect of our study is that the interaction Hamiltonian $%
H_{\text{Int}}$ is in a non-RWA form. Because of the difficulty in studying
this kind of non-RWA interaction, previous studies have used a RWA treatment
for simplicity by assuming the interaction Hamiltonian as
\begin{equation}
H_{\text{Int}}^{\text{RWA}}=\sum_{k}\left( g_{k}\sigma _{+}b_{k}+g_{k}^{\ast
}\sigma _{-}b_{k}^{\dagger }\right) ,  \label{RWA-Hint}
\end{equation}%
where the effect of the counter-rotating terms were omitted. Recently, it
was found that the counter-rotating terms are necessary in order to
accurately describe the spin-boson interaction~\cite{32} and using the
conventional RWA approach may lead to faulty results\thinspace \cite{33}.

Assume that the whole system is initially in the state $\rho _{\text{Tot}%
}\left( 0\right) =\rho _{\text{S}}\left( 0\right) \otimes \rho _{\text{B}}$,
where $\rho _{\text{S}}\left( 0\right) $ is the initial state of the qubit
and chosen as a maximally mixed state $\rho _{\text{S}}\left( 0\right) =%
\mathbb{I}/2$. The bath is considered to be initially in a vacuum state $%
\rho _{\text{B}}=|$vac$\rangle _{\text{BB}}\left\langle \text{vac}%
\right\vert ,$ with $|$vac$\rangle _{\text{B}}\equiv \otimes _{k}|0\rangle
_{k}$. The system-bath coupling spectrum is assumed as a Lorentz-type
\begin{equation}
J\left( \omega \right) =\frac{1}{2\pi }\frac{\gamma \lambda ^{2}}{\left(
\omega -\omega _{0}\right) ^{2}+\lambda ^{2}},  \label{Lorentz}
\end{equation}%
where $\lambda $ is the broadening width of the bosonic mode of frequency $%
\omega $, which is connected to the bath correlation time $\tau _{\text{B}%
}=\lambda ^{-1}$. The relaxation time scale $\tau _{\text{S}}$, on which the
state of the system changes, is related to $\gamma $ by $\tau _{\text{S}%
}=\gamma ^{-1}.$ The $\gamma $ partly reflects the system-bath coupling
strength, because one will obtain the effective coupling strength as $g_{%
\text{eff}}^{2}=\frac{1}{2}\gamma \lambda $ after integrating the spectrum $%
J\left( \omega \right) $ over the entire region of $\omega $.

The evolution under the total Hamiltonian~(\ref{eq:Ham}) can be translated
into the language of quantum channel. Thus the evolution map of the basis
vectors can be described as
\begin{eqnarray}
|g\rangle _{\text{S}}|\text{vac}\rangle _{\text{B}} &\rightarrow &\sqrt{p}%
|g\rangle _{\text{S}}|\text{even}\rangle _{\text{B},g}+\sqrt{1-p}|e\rangle _{%
\text{S}}|\text{odd}\rangle _{\text{B},g},  \notag \\
|e\rangle _{\text{S}}|\text{vac}\rangle _{\text{B}} &\rightarrow &\sqrt{q}%
|e\rangle _{\text{S}}|\text{even}\rangle _{\text{B},e}+\sqrt{1-q}|g\rangle _{%
\text{S}}|\text{odd}\rangle _{\text{B},e},  \notag \\
&&  \label{nonRWA_map}
\end{eqnarray}%
where $|g\rangle _{\text{S}}$ ($|e\rangle _{\text{S}}$) represents the
ground (excited) state of the qubit while the vector $|$even$\rangle _{\text{%
B},i}$ ($|$odd$\rangle _{\text{B},i}$) describes the evolved vector of the
bath, which is actually a superposition of all the number states (e.g., $%
\otimes _{k}|n_{k}\rangle _{k}$) with an even (odd) excitation number (i.e.,
$\sum_{k}n_{k}$ is even or odd). Note that the subscript $i\,$($=e,\,g$) of $%
|$even$\rangle _{\text{B},i}$ ($|$odd$\rangle _{\text{B},i}$) corresponds to
the initial vector $|i\rangle _{\text{S}}|$vac$\rangle _{\text{B}}$. The
vectors satisfy the orthogonality $_{\text{B},i}\langle $even$|$odd$\rangle
_{\text{B},j}=0$ (including $i=j$ and $i\neq j$). They also satisfy the
normalizing condition $_{\text{B},i}\langle $even$|$even$\rangle _{\text{B}%
,i}=1$ and $_{\text{B},i}\langle $odd$|$odd$\rangle _{\text{B},i}=1$. The
overlap $_{\text{B},i}\langle $even$|$even$\rangle _{\text{B},j}$ or $_{%
\text{B},i}\langle $odd$|$odd$\rangle _{\text{B},j}$ ($i\neq j$) outputs a
complex number. The probabilities~$p$ and $q$ are time-dependent, with $%
p\left( t\right) \in \lbrack 0,1]$~and~$q\left( t\right) \in \lbrack 0,1]$.
Then the reduced density matrix of the qubit becomes
\begin{equation}
\rho _{\text{S}}(t)=\left[
\begin{array}{cc}
\rho _{11}\left( t\right)  & \rho _{12}\left( t\right)  \\
\rho _{12}^{\ast }\left( t\right)  & \rho _{22}\left( t\right)
\end{array}%
\right] ,  \label{rho_nonRWA}
\end{equation}%
where $\rho _{11}\left( t\right) =q\rho _{11}\left( 0\right) +(1-p)\rho
_{22}\left( 0\right) $, $\rho _{22}\left( t\right) =(1-q)\rho _{11}\left(
0\right) +p\rho _{22}\left( 0\right) $, and $\rho _{12}\left( t\right) =\rho
_{12}\left( 0\right) \sqrt{pq}Z_{1}\left( t\right) +\rho _{21}\left(
0\right) \sqrt{(1-q)(1-p)}Z_{2}\left( t\right) $. Here $\rho _{i,j}\left(
0\right) $ ($i,j=1,2$) stands for the matrix elements of the qubit's initial
state, while $Z_{1}\left( t\right) \equiv $ $_{\text{B},g}\langle $even$|$%
even$\rangle _{\text{B},e}$ and $Z_{2}\left( t\right) \equiv $ $_{\text{B}%
,e}\langle $odd$|$odd$\rangle _{\text{B},g}$ are the time-dependent complex
numbers.

\textbf{Experimental setup and implementation of the non-RWA channel. }Our
experimental setup is sketched in Fig.\thinspace 1. In Fig.\thinspace 1(a),
pairs of photons with a 810 nm wavelength are produced by pumping a type-I
beta-barium-borate (BBO) crystal with ultraviolet pulses at $405$~nm
centered wavelength. Then one photon is led into a state-preparation
process. That is, the first polarized beam splitter (PBS$_{1}$) selects the
horizontally-polarized state $|H\rangle $ of the photon, and then a half
wave plate (HWP) and a quarter-wave plate (QWP) can rotate $|H\rangle $ into
one of the six eigenstates of the Pauli operators.

Fig.\thinspace 1(b) accomplishes the task of the non-RWA quantum channel. We
use the horizontal and vertical polarization modes $|H\rangle $ and $%
|V\rangle $ to encode the qubit's basis states. The bath acts by a
collective performance of four path modes $|i\rangle _{\text{p}},$ with $%
i\in \{0,1,2,3\}$. In order to briefly introduce the implementation of the
channel, let us start from the output of PBS$_{2}$, where the $H$ and $V$
components are spatially separated so that each one can be rotated with the
wave plate HWP$_{1}$ by angle $\theta _{1}\in \lbrack 0,\pi /4]$ and the
wave plate HWP$_{2}$ by angle $\theta _{2}$ $\in \lbrack 0,\pi /4]$. After
rotating by the wave plate HWP$_{1}$, the $|H\rangle $ mode becomes a
superposition as $|H\rangle \rightarrow \cos 2\theta _{1}|H\rangle +\sin
2\theta _{1}|V\rangle $. By transmitting it along a loop, the superposition
state undergoes PBS$_{2}$ again\ and couples with the spatial modes $%
|0\rangle _{\text{p}}$ and $|1\rangle _{\text{p}}$ encoded as the path
numbers in Fig.~1(b), resulting in the state $\cos 2\theta _{1}|H\rangle
|0\rangle _{\text{p}}+\sin 2\theta _{1}|V\rangle |1\rangle _{\text{p}}$.

It is worth noting that we make use of two Soleil-Babinet compensators (SBC$%
_{1}$ and SBC$_{2}$) in order to append a phase $\phi _{i}=\nu \tau _{i}$ to
the passing components $H$ ($V$). Here, $\nu $ is\ the photon frequency and $%
\tau _{i}=L_{k}n_{l}/c$ is the traveling\ time of the photon across the SBC,
where $L_{k}$ ($k=1,2$) denotes the thickness of SBC$_{1,2}$, $n_{l}$ ($l=H,V
$) indicates the indices of refraction (corresponding to the $H$ and $V$
polarizations), and $c$ is the vacuum speed of light. Hence there are four
phases $\phi _{1,2,3,4}\in \lbrack 0,2\pi ]$ appended to the polarization
states of photons passing SBC$_{1}$ and SBC$_{2}$. This innovative design
enables us to realistically describe the varying phase of the off-diagonal
elements of the system's density matrix.

In Fig.\thinspace 1(b), there are several birefringent calcite beam
displacers (BD$_{1,2,3,4}$) which deviate the $H$ component and transmit the
$V$ one. Among them, we insert some wave plates to implement operations on
the polarization states. The angles of HWP$_{3,4}$ are adjusted as $\theta
_{3}\in \lbrack 0,\pi /4]$ and $\theta _{4}\in \lbrack 0,\pi /4]$ to
transform a single component ($H$ or $V$) into a superposition form, while
the angles of HWP$_{5,6,7,8}$ are fixed at $\theta _{5,6,7,8}=\pi /4$ to
convert $H$ into $V$ or vice versa. The input-output states of
Fig.\thinspace 1(b) are mapped as follows(see details in the method section):

\begin{eqnarray}
|H\rangle |0\rangle _{\text{p}} &\rightarrow &\cos 2\theta _{1}e^{i\phi
_{1}}|H\rangle |0\rangle _{\text{p}}+\sin 2\theta _{1}e^{i\phi
_{2}}|V\rangle \left\vert \Psi \right\rangle _{1,3},  \notag \\
|V\rangle |0\rangle _{\text{p}} &\rightarrow &\cos 2\theta _{2}e^{i\phi
_{3}}|V\rangle \left\vert \Psi \right\rangle _{0,2}-\sin 2\theta
_{2}e^{i\phi _{4}}|H\rangle |3\rangle _{\text{p}},  \notag \\
&&  \label{HV_map}
\end{eqnarray}%
where $\left\vert \Psi \right\rangle _{1,3}\equiv \cos 2\theta _{3}|1\rangle
_{\text{p}}-\sin 2\theta _{3}|3\rangle _{\text{p}}$ and $\left\vert \Psi
\right\rangle _{0,2}\equiv \cos 2\theta _{4}|2\rangle _{\text{p}}-\sin
2\theta _{4}|0\rangle _{\text{p}}$. By comparing Eq.\thinspace (\ref{HV_map}%
) with Eq.\thinspace (\ref{nonRWA_map}), it can be seen that the reduced
density operator of the system in Eq.\thinspace (\ref{rho_nonRWA}) is
successfully reproduced, by setting the parameters $q=\cos ^{2}2\theta _{1}$%
, $p=\cos ^{2}2\theta _{2}$, $Z_{1}=\sin \left( 2\theta _{4}\right)
e^{i(\phi _{3}-\phi _{1})}$, and $Z_{2}=\sin \left( 2\theta _{3}\right)
e^{i(\phi _{2}-\phi _{4})}$. It is noted that according to the hierarchy
equation method\thinspace \cite{25,29}, the qubit state at an arbitrary time
$t$ can be simulated by adjusting the parameters $\theta _{i}$ and $\phi
_{i} $ according to the theoretical results obtained by the hierarchy
equation method.

In Fig.\thinspace 1(c), the density matrix of the output state is
reconstructed by quantum tomography process where ten different coincidence
measurement bases are set by QWPs, HWPs and PBSs. Eight of the bases are set
along paths 0 and 3, while the rest are set along the paths 1 and 2.
Finally, the photons are detected by single-photon detectors equipped with
10 nm interference filters.

As an application, our proposed experimental setup (Fig.~1) can be used to
implement the channel governed by the RWA treatment of the interaction
Hamiltonian in Eq.\thinspace (\ref{RWA-Hint}). Based on the evolution map in
Eq.\thinspace (\ref{HV_map}), one can set the HWP angles $\theta _{1}=0$, $%
\theta _{4}=\pi /4$, and adjust $\theta _{2}\in \lbrack 0,\pi /4]$ according
to the theoretical results with RWA. Furthermore, the phases $\phi _{1,3} $
is adjusted in $[0,2\pi ]$ according to the theoretical results, and $\phi
_{2,4}$ is set randomly. Then the evolution map in the Schr\"{o}dinger
picture reads:
\begin{eqnarray}
&&~|H\rangle |0\rangle _{\text{p}}\rightarrow e^{i\phi _{1}}|H\rangle
|0\rangle _{\text{p}},  \notag \\
&&~|V\rangle |0\rangle _{\text{p}}\rightarrow \sqrt{P_{\text{AD}}}e^{i\phi
_{3}}|V\rangle |0\rangle _{\text{p}}+\sqrt{1-P_{\text{AD}}}e^{i\phi
_{4}}|H\rangle |3\rangle _{\text{p}},  \notag \\
&&  \label{RWA_channel}
\end{eqnarray}%
where $P_{\text{AD}}=\cos ^{2}(2\theta _{2})$, and the spacial modes $%
|0\rangle _{\text{p}}$ and $|3\rangle _{\text{p}}$ correspond to the paths $%
0 $ and $3$ in Fig.~1(b), respectively. If we further set the phases $\phi
_{1,3,4}=0$ according to the interaction picture, Eq.\thinspace (\ref%
{RWA_channel}) describes the so-called amplitude decay channel which is just
the channel implemented in Ref.~\cite{34,35}. The authors in~\cite{34}
tested strong coupling cases in order to explore abundant non-Markovianity.
However, it was theoretically predicated that the channels with or without
RWA can cause quite different non-Markovian behavior especially in
strong-coupling cases~\cite{36}. From this point of view, considering the
quantum channels without RWA is of importance and an interesting question.

Another well-known channel, the so-called phase damping channel considered
in\thinspace \cite{35}, is also easy to implement in our scheme. Based on
Eq.~(\ref{HV_map}), by setting $\theta _{1,2}=0$, $\phi _{1,2,3,4}=0$ and
adjusting $\theta _{4}\in \lbrack 0,\pi /4]$, one can have the following
evolution map:
\begin{eqnarray}
&&~|H\rangle |0\rangle _{\text{p}}\rightarrow |H\rangle |0\rangle _{\text{p}%
},  \notag \\
&&~|V\rangle |0\rangle _{\text{p}}\rightarrow \sqrt{P_{\text{PD}}}|V\rangle
|0\rangle _{\text{p}}-\sqrt{1-P_{\text{PD}}}|V\rangle |2\rangle _{\text{p}},
\end{eqnarray}%
where $P_{\text{PD}}=\sin ^{2}(2\theta _{4})$, and the spacial modes $%
|0\rangle _{\text{p}}$ and $|2\rangle _{\text{p}}$ correspond to the paths $%
0 $ and $2$ in Fig.~1(b), respectively.

\textbf{\ TS parameter. }In the TS problem,\textbf{\ }a system is sent to a
distant receiver (say Bob) through a quantum channel. Before Bob receives
the system, a detector or manipulator (say Alice) performs some measurements
on the system. Then the TS problem refers to the characterization of the
influence of Alice's measurement at an initial time $t_{A}$ (let $t_{A}=0$
in this paper) on the final state captured by Bob at a later time $t_{B}$.
The TS in the qubit systems can be detected by\textbf{\ }a concept named as
TS parameter $S_{N},$\ which is defined in terms of a temporal analogue of
the steering inequality\thinspace \cite{21,21-1},
\begin{equation}
S_{N}\equiv \sum\limits_{i=1}^{N}E\left( \left\langle
B_{i,t_{B}}\right\rangle _{A_{i,t_{A}}}^{2}\right) \leq 1,
\label{S_inequality}
\end{equation}%
where $A_{i,t_{A}}$ ($B_{i,t_{B}}$) stands for the $i$th observable measured
by Alice (Bob) at $t_{A}$ ($t_{B}$) and the number of observable is $N=2$ or
$3$.
\begin{equation}
E\left( \left\langle B_{i,t_{B}}\right\rangle _{A_{i,t_{A}}}^{2}\right)
\equiv \sum\limits_{a=\pm 1}P\left( a|A_{i,t_{A}}\right) \left\langle
B_{i,t_{B}}\right\rangle _{A_{i,t_{A}}=a}^{2},
\end{equation}%
with $P\left( a|A_{i,t_{A}}\right) $ being the probability of Alice's
measurement outcome $a=+1$ or $-1$.\ The Bob's expectation value,
conditioned on Alice's measurement outcome, is defined as%
\begin{equation}
\left\langle B_{i,t_{B}}\right\rangle _{A_{i,t_{A}}=a}\equiv
\sum\limits_{b=\pm 1}bP\left( B_{i,t_{B}}=b|A_{i,t_{A}}=a\right) ,
\end{equation}%
where $P\left( B_{i,t_{B}}=b|A_{i,t_{A}}=a\right) $ denotes the condition
probability of Bob's measurement outcome $b$ (at $t_{B}$) on the evolved
state starting from the collapsed version after Alice's measurement with the
outcome $a$ (at $t_{A}$). In this paper, we take the case of $N=2$ into
account. The violation of the inequality in Eq.\thinspace (\ref{S_inequality}%
), i.e. $S_{2}>1$, is a\ sufficient condition for the steerability.
Therefore, during an evolution, one can define the steerable durations
conditioned by $S_{2}>1$.

\textbf{Experimental and numerical results of TS parameter.} We consider
that Alice chooses a pair of the Pauli operators $\left\{ \sigma _{i\text{ }%
},\sigma _{j}\right\} $ $(i,j=x,y,z)$ as the observables ($A_{i}$) measured\
on the initial state $\rho _{\text{S}}=\mathbb{I}/2$ of the qubit. After the
measurement, the qubit state collapses to one of the six eigenstates of the
Pauli operators with a probability $P\left( a|A_{i,t_{A}}\right) =1/2$. This
process is usually difficult to implement in experiments, since it requires
a set of nondestructive measurements. An equivalent way, adopted in this
experiment, is to assume that Alice prepares qubit states by rotating the
polarization mode $\left\vert H\right\rangle $ into one of the six
eigenstates of the Pauli operators, and correspondingly multiplies a
probability of $P\left( a|A_{i,t_{A}}\right) $. This preparation is
completed by sequentially using the PBS$_{1}$, a HWP, and a QWP [see
Fig.\thinspace 1(a)]. Then the qubit in the prepared state is sent through
the quantum channel [simulated in Fig.\thinspace 1(b)] to Bob, who performs
tomography measurements [Fig.\thinspace 1(c)]. Therefore, Bob obtains the
condition probabilities $P\left( B_{i,t_{B}}=b|A_{i,t_{A}}=a\right) $ and
calculates $S_{2}$.\ Theoretically speaking, $S_{2}$ is a function of the
time $t$, the channel parameter $\gamma $, and the parameter $\lambda $.
Actually, in our experiment, the dynamics of $S_{2}$\ is simulated by
adjusting the angles of HWPs and the thicknesses of SBCs. The experimental
errors are estimated from the statistical variation of photon counts, which
satisfy the Poisson distribution.

Figure\thinspace 2 shows $S_{2}$ vs evolution time scaled by $\omega _{0}$.
In Fig.\thinspace 2(a) and Fig.\thinspace 2(b), the measurement bases are $%
\left\vert +\right\rangle $ ($\left\vert -\right\rangle $) and $\left\vert
0\right\rangle $ ($\left\vert 1\right\rangle $), which correspond to the
eigenstates of $\sigma _{x\text{ }}$and $\sigma _{z}$, respectively. Our
experimental and theoretical results are in good agreement and show the
oscillation of TS parameter $S_{2}$ with time. More important is the
steering limit, i.e., $S_{2}=1$, which is marked by a red-dashed horizontal
line in Fig.~2. Above this limit, steerability is valid. The vertical dashed
lines highlight the steerable durations corresponding to $S_{2}>1$. For the
sake of comparison, we study two kinds of channels, i.e., the non-RWA
channel and the RWA channel. The former is modeled by the
Hamiltonian\thinspace (\ref{non-RWA-Hint}) and experimentally simulated
according to the evolution map in Eq.\thinspace (\ref{HV_map}), while the
latter is modeled by the Hamiltonian\thinspace (\ref{RWA-Hint}) and
experimentally simulated based on the evolution map in Eq.\thinspace (\ref%
{RWA_channel}). In both of the cases of RWA and non-RWA channels, $S_{2}$ is
suppresed below the steering limit in most of the evolution periods due to
the quantum decoherence effects. However, the difference between the two
cases is\ obvious. There are more peaks of $S_{2}$ over the steering limit
in the RWA case [Fig.\thinspace 2(b)] than the non-RWA case [Fig.\thinspace
2(a)], which implies that the RWA channel appears to provide longer
steerable durations. Similar conclusions can be made by comparing the
results in Fig.~2(c) and Fig.~2(d), where another set of measurement bases
are chosen, i.e., $\left\vert +\right\rangle $ ($\left\vert -\right\rangle $%
) and $\left\vert R\right\rangle $ ($\left\vert L\right\rangle $)
(eigenstates of $\sigma _{x}$ and $\sigma _{y}$, respectively). However we
should point out that the extra steerable durations in RWA cases are
inauthentic, due to the essential defects in characterizing the system-bath
interaction by using RWA.

\textbf{Experimental and numerical results of TS weight. }We also
experimentally test the steerable weight of TS, i.e., $W_{\text{TS}}$ (see
the definition in the method section), as illustrated in Fig.\thinspace 3.
The experimental implementations in the input-state preparation (at Alice's
side) and the tomography measurement on the output states (at Bob's side)
are the same as those in Fig.\thinspace 2. The non-RWA case [Fig.\thinspace
3(a)] and RWA case [Fig.\thinspace 3(b)] are investigated. The parameters $%
\gamma $ and $\lambda $ are chosen as same as those in Fig.\thinspace 2.
Consequently, we compare the results shown in Figs.\thinspace 3(a) and (b)
with those in Figs.\thinspace 2(c) and (d). Since $W_{\text{TS}}$ is defined
according to the sufficient and necessary condition of the exsitence of TS,
the data of $W_{\text{TS}}$ precisely tells us when the TS exists and
disappears, especially for the durations below the TS limit $S_{2}=1$
[Fig.\thinspace 2(c) and (d)], where the criterion $S_{2}$ is disabled to
detect TS.

By comparison of Fig.\thinspace 3(a) with Fig.\thinspace 3(b), more
interesting phenomena are found, i.e., there are obvious \textquotedblleft
sudden death\textquotedblright\ and \textquotedblleft
revival\textquotedblright\ of TS in the non-RWA channel, whereas they never
appear in the RWA channel where TS tends to zero asymptotically. We shall
emphasize that the quantum correlation like TS inevitably undergoes a sudden
change to zero rather than a gradual decrease, especially when the
characterization of the system-bath interaction becomes close to the actual
situation. This also reminds us of the previous famous report on the sudden
death of entanglement\thinspace \cite{39}.

\section*{Discussion}

With the proposed setup, we have experimentally implemented a non-RWA
quantum channel and simulated the dynamics of a qubit system interacting
with a bosonic bath, without applying Born, Markov and rotating-wave
approximations. This kind of quantum channel provides a more realistic
description of the environmental impact and allows a variety of
investigation on coherence dynamics in open systems. Based on this channel,
we have studied the important TS problem experimentally, which is attracting
much attention recently. By means of the TS inequality, the experimental
data agree well with the theoretical results, showing that quantum
decoherence can significantly shorten the steerable durations. Furthermore,
our investigation shows that although the RWA channel seems to provide
longer steerable durations than the non-RWA case, the results without\ RWA
are closer to realistic situations and thus more reliable.

By investigating the TS weight, we observed some new interesting phenomena
in the non-RWA channel, i.e., the \textquotedblleft sudden
death\textquotedblright\ and \textquotedblleft revival\textquotedblright\ of
the TS, which however do not appear in the RWA channel. The RWA channel
presents a false superiority compared with the non-RWA case.

Characterizing the quantum channel as precisely as possible is necessary and
important in the study of quantum steering dynamics. Our studies show that
an often-used approximation method such as the RWA leads to a deviation from
the facts. It remains an open question to test other quantum correlation
dynamics in the non-RWA quantum channel. On the other hand, since quantum
steering inequalities are closely linked with the security of the quantum
cryptographic protocols, our findings are expected to have valuable
applications in secure quantum communications.

\section*{Methods}

\textbf{Implementation of the evolution map in Eq.\thinspace (\ref{HV_map}).
}The input-output states for the setup in Fig.\thinspace 1(b) are shown as
follows:
\begin{eqnarray}
|H\rangle &\overset{\text{PBS}_{2},\text{HWP}_{1}}{\longrightarrow }&\cos
2\theta _{1}|H\rangle |2\rangle +\sin 2\theta _{1}|V\rangle |1\rangle
\overset{\text{SBC}_{1,2}}{\longrightarrow }\cos 2\theta _{1}e^{i\phi
_{1}}|H\rangle |2\rangle +\sin 2\theta _{1}e^{i\phi _{2}}|V\rangle |1\rangle
\notag \\
&\overset{\text{BD}_{1,2}}{\longrightarrow }&\cos 2\theta _{1}e^{i\phi
_{1}}|H\rangle |0\rangle +\sin 2\theta _{1}e^{i\phi _{2}}|V\rangle |1\rangle
\notag \\
&\overset{\text{HWP}_{5,3}}{\longrightarrow }&\cos 2\theta _{1}e^{i\phi
_{1}}|V\rangle |0\rangle +\sin 2\theta _{1}e^{i\phi _{2}}(\cos 2\theta
_{3}|V\rangle -\sin 2\theta _{3}|H\rangle )|1\rangle  \notag \\
&\overset{\text{BD}_{3,4}}{\longrightarrow }&\cos 2\theta _{1}e^{i\phi
_{1}}|V\rangle |0\rangle +\sin 2\theta _{1}e^{i\phi _{2}}(\cos 2\theta
_{3}|V\rangle |1\rangle -\sin 2\theta _{3}|H\rangle |3\rangle )  \notag \\
&\overset{\text{HWP}_{7,8}}{\longrightarrow }&\cos 2\theta _{1}e^{i\phi
_{1}}|H\rangle |0\rangle +\sin 2\theta _{1}e^{i\phi _{2}}(\cos 2\theta
_{3}|V\rangle |1\rangle -\sin 2\theta _{3}|V\rangle |3\rangle ),~~~~~~~~~~
\end{eqnarray}%
\begin{eqnarray}
|V\rangle &\overset{\text{PBS}_{2},\text{HWP}_{2}}{\longrightarrow }&\cos
2\theta _{2}|V\rangle |2\rangle -\sin 2\theta _{2}|H\rangle |1\rangle
\overset{\text{SBC}_{1,2}}{\longrightarrow }\cos 2\theta _{2}e^{i\phi
_{3}}|V\rangle |2\rangle -\sin 2\theta _{2}e^{i\phi _{4}}|H\rangle |1\rangle
\notag \\
&\overset{\text{BD}_{1,2}}{\longrightarrow }&\cos 2\theta _{2}e^{i\phi
_{3}}|V\rangle |2\rangle -\sin 2\theta _{2}e^{i\phi _{4}}|H\rangle |3\rangle
\notag \\
&\overset{\text{HWP}_{6,4}}{\longrightarrow }&\cos 2\theta _{2}e^{i\phi
_{3}}(\cos 2\theta _{4}|V\rangle -\sin 2\theta _{4}|H\rangle )|2\rangle
-\sin 2\theta _{2}e^{i\phi _{4}}|V\rangle |3\rangle  \notag \\
&\overset{\text{BD}_{3,4}}{\longrightarrow }&\cos 2\theta _{2}e^{i\phi
_{3}}(\cos 2\theta _{4}|V\rangle |2\rangle -\sin 2\theta _{4}|H\rangle
|0\rangle )-\sin 2\theta _{2}e^{i\phi _{4}}|V\rangle |3\rangle  \notag \\
&\overset{\text{HWP}_{7,8}}{\longrightarrow }&-\sin 2\theta _{2}e^{i\phi
_{4}}|H\rangle |3\rangle +\cos 2\theta _{2}e^{i\phi _{3}}(\cos 2\theta
_{4}|V\rangle |2\rangle -\sin 2\theta _{4}|V\rangle |0\rangle
),~~~~~~~~~~\qquad
\end{eqnarray}%
where the annotations above the arrows represent the devices through which
the operations on the states are performed.

\textbf{TS weight.} Alice measures the observable $A_{i}$ on the system's
state at an initial time $t_{A}$, and gets the outcome $a$ with a
probability of $P\left( a|A_{i,t_{A}}\right) $. Assume that there are $N$
observables, i.e. $A_{i}$ with $i=1,...,N$, and each of them is $m$
dimension (the case of $m=2$ is considered in this paper). After the
measurement, the system's state is mapped to $\rho _{a|A_{i}}$. Then, the
system is sent to Bob through a quantum channel $\Lambda $. At time $t_{B}$,
Bob receives the system and performs tomography measurements to obtain the
state $\varsigma _{a|A_{i}}=\Lambda \left( \rho _{a|A_{i}}\right) $.\textbf{%
\ }In order to precisely quantify TS, a concept named TS weight, i.e. $W_{%
\text{TS}}$, is introduced via a semidefinite program as\thinspace \cite%
{17,22}:
\begin{equation}
W_{\text{TS}}\equiv 1-\max \text{ Tr}\sum\limits_{\lambda }\varrho _{\lambda
},  \label{TSW_1}
\end{equation}%
subject to
\begin{equation}
\tilde{\varsigma}_{a|A_{i}}-\sum\limits_{\lambda }D_{\lambda }\left(
a|A_{i}\right) \varrho _{\lambda }\geq 0,\text{ \ }\forall a,A_{i},
\label{TSW_2}
\end{equation}%
and
\begin{equation}
\varrho _{\lambda }\geq 0,  \label{TSW3}
\end{equation}%
where $\tilde{\varsigma}_{a|A_{i}}\equiv P\left( a|A_{i,t_{A}}\right)
\varsigma _{a|A_{i}}$ stands for the un-normalized states received by Bob.
The task of Bob is to check whether the states he receives can be written in
a hidden-state form, i.e.\thinspace $\sum_{\lambda }D_{\lambda }\left(
a|A_{i}\right) \varrho _{\lambda }$ in Eq.\thinspace (\ref{TSW_2}), where $%
\lambda $ ($=1,...,m^{N}$) indicates a classical random variable. $\varrho
_{\lambda }$ indicates a set of positive semidefinite matrices held by Bob,
and $D_{\lambda }\left( a|A_{i}\right) $ is the deterministic single-party
conditional probability according to Alice's measurement outcome \cite{17,22}%
. Nonzero $W_{\text{TS}}$ implies that Bob cannot classically fabricate
Alice's measurement results, and thus the quantum temporal correlation
(i.e., the TS) exists between Alice and Bob. Note that $W_{\text{TS}}$ comes
from a sufficient and necessary characterization of steerability and
quantifies the TS precisely.

\begin{addendum}

\item[Acknowledgments]
Z.S. is supported by the National Natural Science Foundation of China under
Grants No. 11375003, the Zhejiang Natural Science Foundation under Grant No.
LY17A050003, the Program for HNUEYT under Grant No. 2011-01-011. C.P.Y is
supported in part by the Ministry of Science and Technology of China under
Grant No. 2016YFA0301802, the National Natural Science Foundation of China
under Grant Nos. [11074062, 11374083]. Z.S., X.Q.X, J.S.J, and C.P.Y. are also supported by
the Zhejiang Natural Science Foundation under Grant No. LZ13A040002 and the
funds from Hangzhou City for the Hangzhou-City Quantum Information and
Quantum Optics Innovation Research Team. The authors thank Prof.~Chuan-Feng Li for valuable suggestions and Dr.~Kai Sun and Dr.~Xiaoming Hu for helpful discussions in the experimental implementation.

\item[Author contributions]
S.J.X. and Z.S. provided the idea. S.J.X., Y.Z., Z.S., and C.P.Y. designed the experiment. S.J.X., Y.Z., and Z.S. conducted the experiment and analyzed the data. Z.S., L.Y., X.Q.X., J.M.L., K.F.C., and C.P.Y. developed the theory. Z.S. and J.S.J. finished the numerical simulation. The manuscript was jointly written by the authors.
\item[Additional information]
\item[Competing financial interests:]The authors declare no competing
financial interests.
\end{addendum}

\clearpage

\textbf{Figure 1:}~Experimental setup and the stages of the experiment.
\textbf{(a)} The photon pairs with a 810\thinspace nm wavelength are
produced via spontaneous parametric down-conversion. One of the two photons
is used as the trigger for the coincident counts. The other photon is led to
the preparation unit consisting of a polarized beam splitter (PBS$_{1}$), a
half wave plate (HWP), and a quarter-wave plate (QWP). In the TS problems,
this photon is prepared into one of the six eigenstates of the Pauli
operators $\sigma _{x,y,z}$. \textbf{(b)} Simulation of the quantum channel
without RWA in Eq.\thinspace (\ref{nonRWA_map}). The angles of HWP$%
_{1,2,3,4} $ are adjusted in $[0,\pi /4]$, while the angles of HWP$%
_{5,6,7,8} $ are set at $\pi /4$. Two Soleil-Babinet compensators (SBC$_{1}$
and SBC$_{2}$) add relative phases to the passing components $H$ and $V$,
respectively. The birefringent calcite beam displacers (BD$_{1,2,3,4}$)
couple the polarization states $\left\vert H\right\rangle $ and $\left\vert
V\right\rangle $ with the spacial modes $\left\vert i\right\rangle _{\text{p}%
}$ ($i=0,\,1,\,2,\,3$). \textbf{(c)} Quantum state tomography is implemented
by two QWPs, four HWPs, and two PBSs. Finally, two single-photon detectors
equipped with two 10\thinspace nm interference filters (IFs) are used for
the photon counting.

\textbf{Figure 2:} TS parameter $S_{2}$ versus scaled time $\omega _{0}t$ in
the non-RWA and RWA channels. (a) and (b) correspond to the measuring bases $%
|+\rangle ~(|-\rangle )$ and $|0\rangle ~(|1\rangle )$ which are the
eigenstates of $\sigma _{x}$ and $\sigma _{z}$, respectively. While, (c) and
(d) correspond to the measuring bases $|+\rangle ~(|-\rangle )$ and $%
|R\rangle ~(|L\rangle )$ which are the eigenstates of $\sigma _{x}$ and $%
\sigma _{y}$, respectively. The channel parameters, i.e., the system-bath
coupling parameters $\gamma =2.5\omega _{0}$ and the broadening width of the
bath mode $\lambda =0.05\omega _{0}$, which result in an effective strength
of the system-bath coupling, i.e., $g_{\text{ eff}}=0.25\omega _{0}$.
Horizontal red dashed lines indicate the steering limit. Vertical dashed
lines point out the steerable durations corresponding to $S_{2}>1$. Inset:
enlarged drawing with more data for the peaks of $S_{2}$ close to or beyond
the steering limit.

\textbf{Figure 3:} TS weight vs scaled time $\omega _{0}t$ in the non-RWA
and RWA channels. The measuring bases in (a) and (b) are $|+\rangle
~(|-\rangle )$ and $|0\rangle ~(|1\rangle )$ which are the eigenstates of $%
\sigma _{x}$ and $\sigma _{z}$, respectively. The values of parameters $%
\gamma $ and $\lambda $ are chosen as same as those in Fig.\thinspace 2.
\clearpage

\begin{figure*}[tbp]
\begin{center}
\epsfig{file=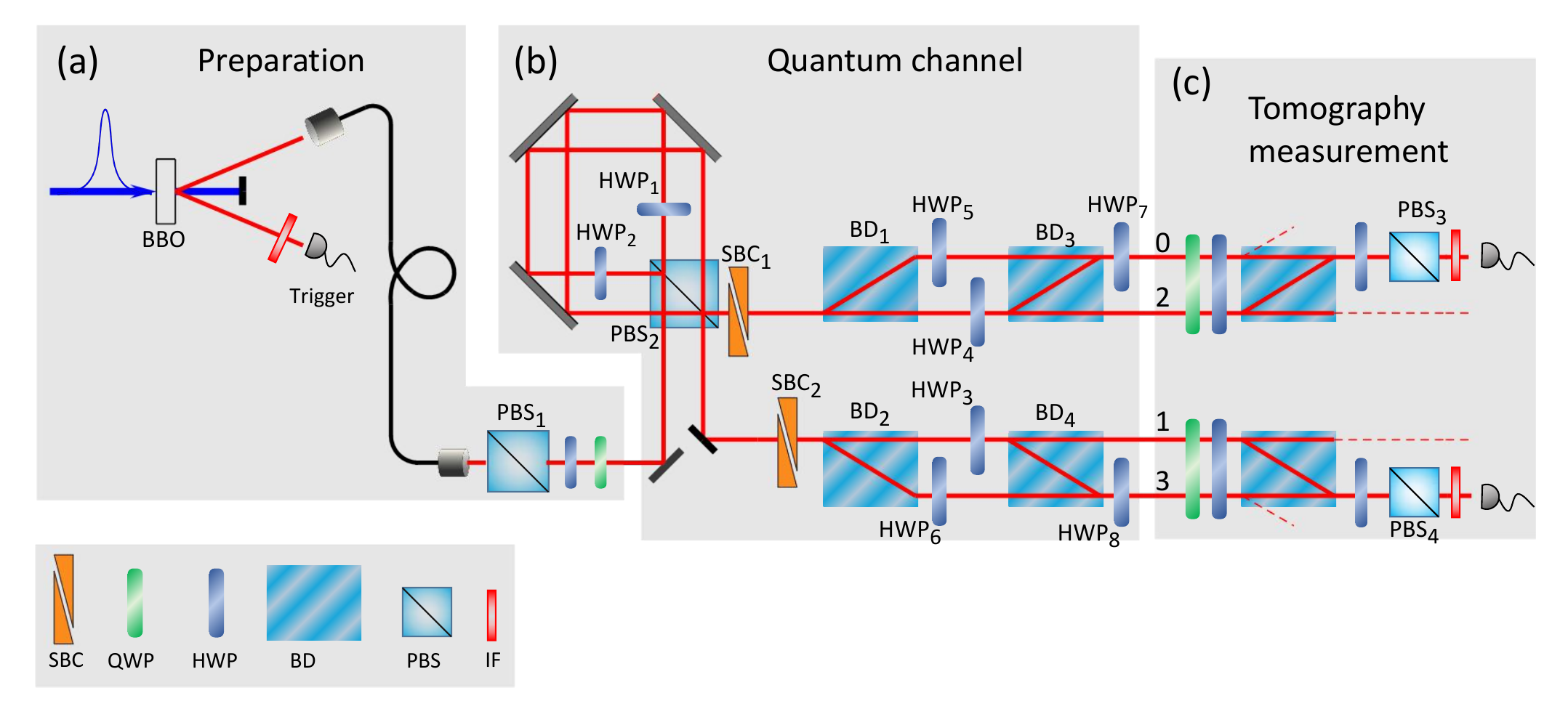,width=17cm, height=8cm}
\end{center}
\caption{}
\label{fig:1}
\end{figure*}

\begin{figure*}[tbp]
\begin{center}
\epsfig{file=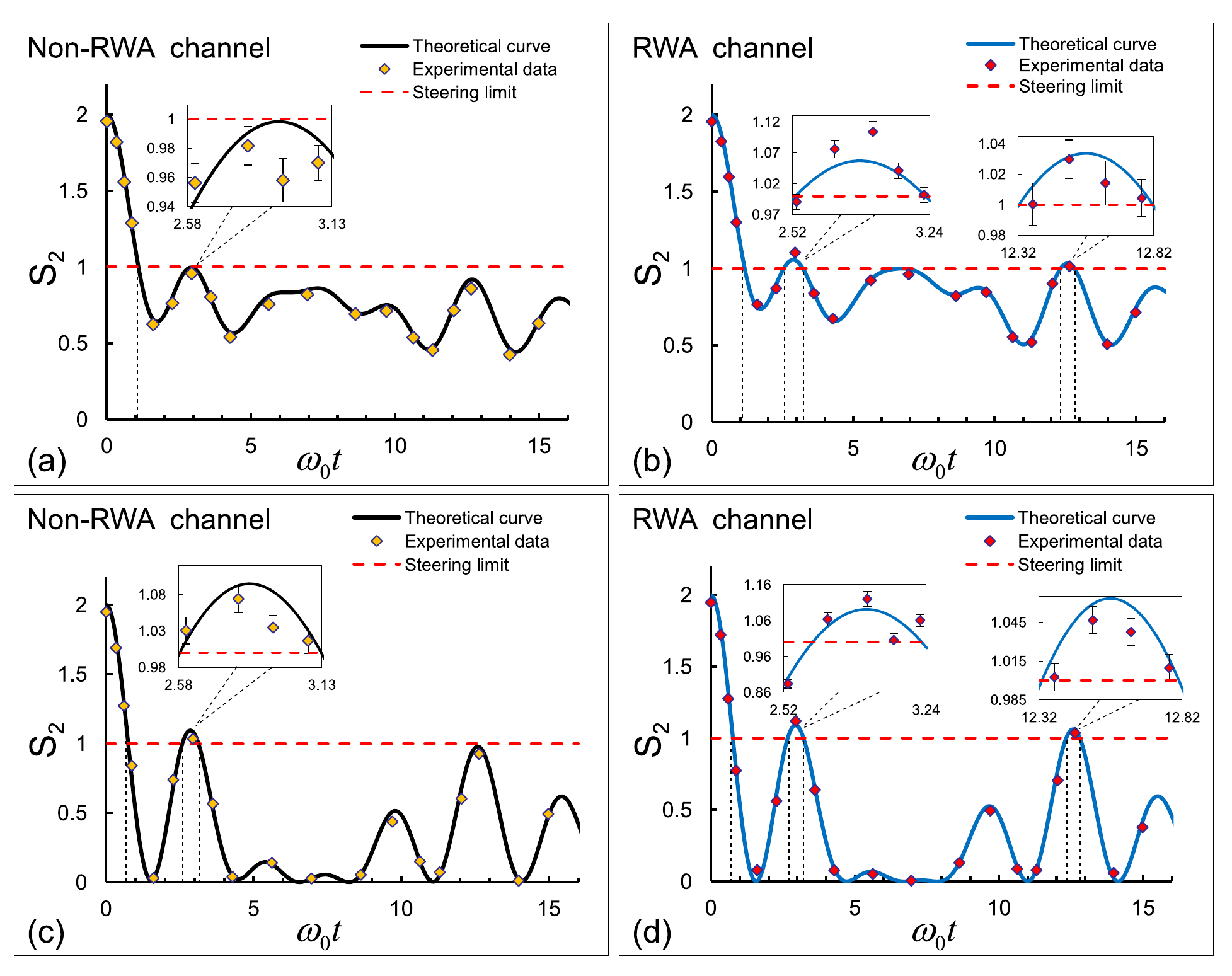,width=17cm, height=12cm}
\end{center}
\caption{}
\label{fig:2}
\end{figure*}

\begin{figure*}[tbp]
\begin{center}
\epsfig{file=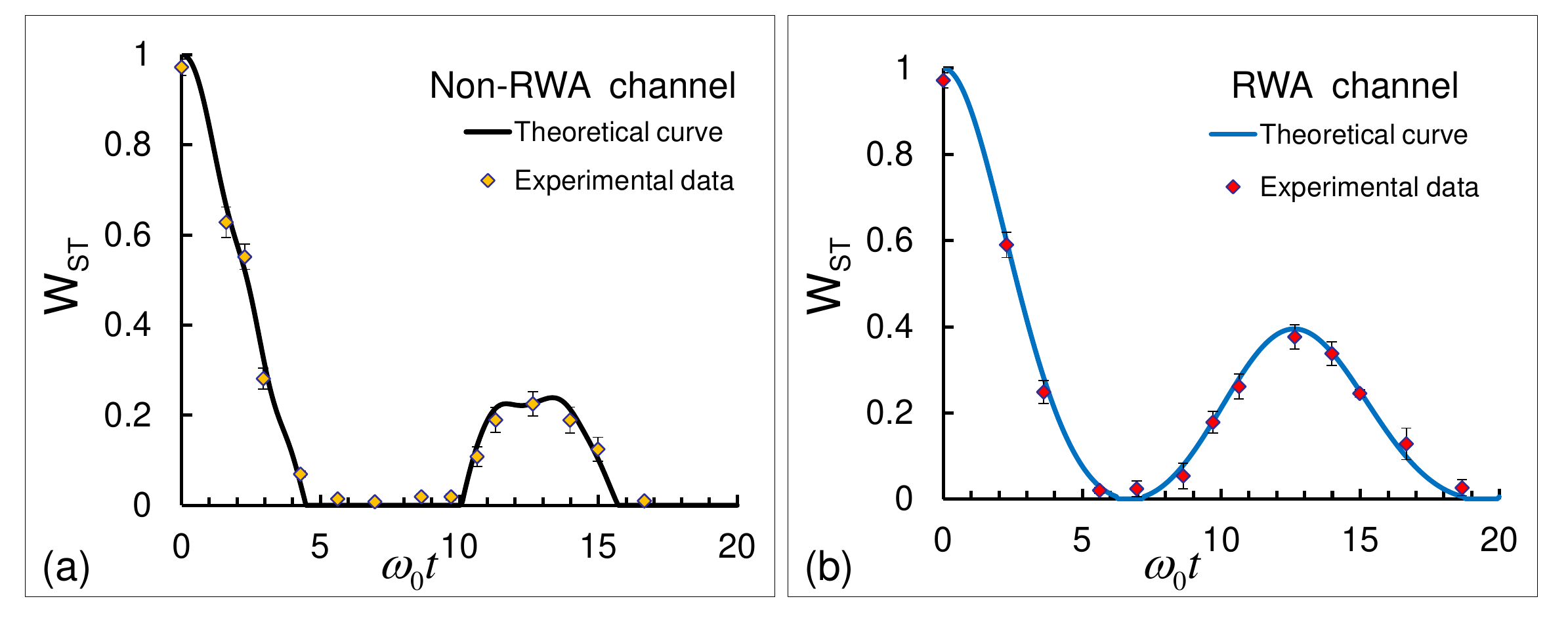,width=17cm, height=7cm}
\end{center}
\caption{}
\label{fig:3}
\end{figure*}

%\begin{figure}
%\begin{center}
%\epsfig{file=Fig-S1.eps,width=12cm}
%\end{center}\caption{}
%\label{Schematic diagram of quantum multiple access network with
%chaotic phase shifters}
%\end{figure}

\end{document}